\documentclass[a4paper]{jpconf}

\usepackage{graphicx}
\usepackage{amsmath,amssymb}

\begin{document}

\title{Knotting and higher order linking in physical systems}

%\author{Roman V. Buniy$^1$, Martha J. Holmes$^2$ and
%Thomas W. Kephart$^{2,}$\footnote{\hspace{-.07in}$^{^*}$ speaker at Quantum Theory and Symmetries 6, University of Kentucky, July (2009)}$^{^*}$ } 

\author{Roman V. Buniy$^1$, Martha J. Holmes$^2$ and Thomas
  W. Kephart$^{2,}$\footnote[3]{speaker at Quantum Theory and
    Symmetries 6, University of Kentucky, July 2009}}

\address{$^1$ School of Earth and Space Exploration, and Physics
  Department, Arizona State University, Tempe, AZ 85287, USA \\ $^2$
  Department of Physics and Astronomy, Vanderbilt University,
  Nashville, TN 37235, USA}

\ead{ roman.buniy@gmail.com}
\ead{ martha.j.holmes@vanderbilt.edu}
\ead{ tom.kephart@gmail.com}

\begin{abstract}
We discuss physical systems with topologies more complicated than
simple gaussian linking. Our examples of these higher topologies are
in non-relativistic quantum mechanics and in QCD.
\end{abstract}

\section{Introduction}
Knots occur in nature. Our focus will be on tight knots and links, and
our main examples will come from quantum mechanics and quantum field
theory, but we will begin with a mention of knots in plasma physics
and biology, since these two areas have lead the way in applications
of knot theory to physical systems \cite{ASetal}.

\section{Examples from classical physics}

In plasma physics, magnetic flux tubes are under tension and tend to
contract. One can observe plasma phenomena directly via images from
the SOHO satellite where Sun spots and solar flares are clearly
visible. The flares reach millions of miles above the surface of the
sun. Pieces break off and are carried by the solar wind into the wider
solar system. These free pieces of plasma have embedded magnetic
fields, and some have non-vanishing magnetic helicity. As they
interact with planet magnetospheres, they can exchange energy and
magnetic helicity with them. The remainder is carried off to
interstellar space.  At a larger scale, our Milky Way and other
galaxies have galactic winds due to all the stars they contain. Some
of this plasma rushes off into intergalactic space. Hence it is
thought that our Universe is full of plasma with trapped magnetic
field, which in general has non-vanishing magnetic helicity.

\section{Knots in biosystems}

A plasma tends to expand when released into the vacuum, but if there
are magnetic fields they can resist this expansion to some degree. In
cases where there is magnetic helicity the expansion can be halted
when an quasi-equilibrium is reached. In an ideal plasma magnetic
helicity is a conserved quantity.  There are minimal energy
configurations determined by topology after minimization of the energy
holding magnetic helicity fixed. These configurations are known as
Taylor states \cite{Woltjer,Moffatt,Taylor}, and undisturbed plasma
regions would tend to these states.

Tight knots were first discussed and some of their lengths estimated
in \cite{ASetal}. However, most biosystems that have knotting do not
have tight knots.  DNA knots are a case in point. Many knot types have
been discovered in DNA, but none are tight due to the persistence
length of the DNA strand.  Proteins can also be knotted, e.g., a
computer model of the knotted protein in methanobacterium
thermoautotrophicum has been studied by a group at Argonne National
Laboratory. Even though proteins are compact objects, as knots they do
not qualify as tight either.

\section{Examples from quantum physics}
Another example of a physical systems with linking and knotting is
 the Aharonov-Bohm effect~\cite{Aharonov:1959fk} and its generalizations.
 The magnetic Aharonov-Bohm effect, see Fig.~1,
results when a charged particle travels around a closed path in a
region of vanishing magnetic field but non-vanishing vector potential.
The wave function of the particle is affected by the vector potential
and an interference pattern proportional to
magnetic flux occurs at a detection screen. The conclusion one draws
is that the vector potential is more fundamental than the magnetic
field. Definitive experiments were done in the mid 1980s~\cite{Tomoyama}.

\begin{figure}[h]
\begin{center}
\begin{minipage}{14pc}
\includegraphics[width=14pc]{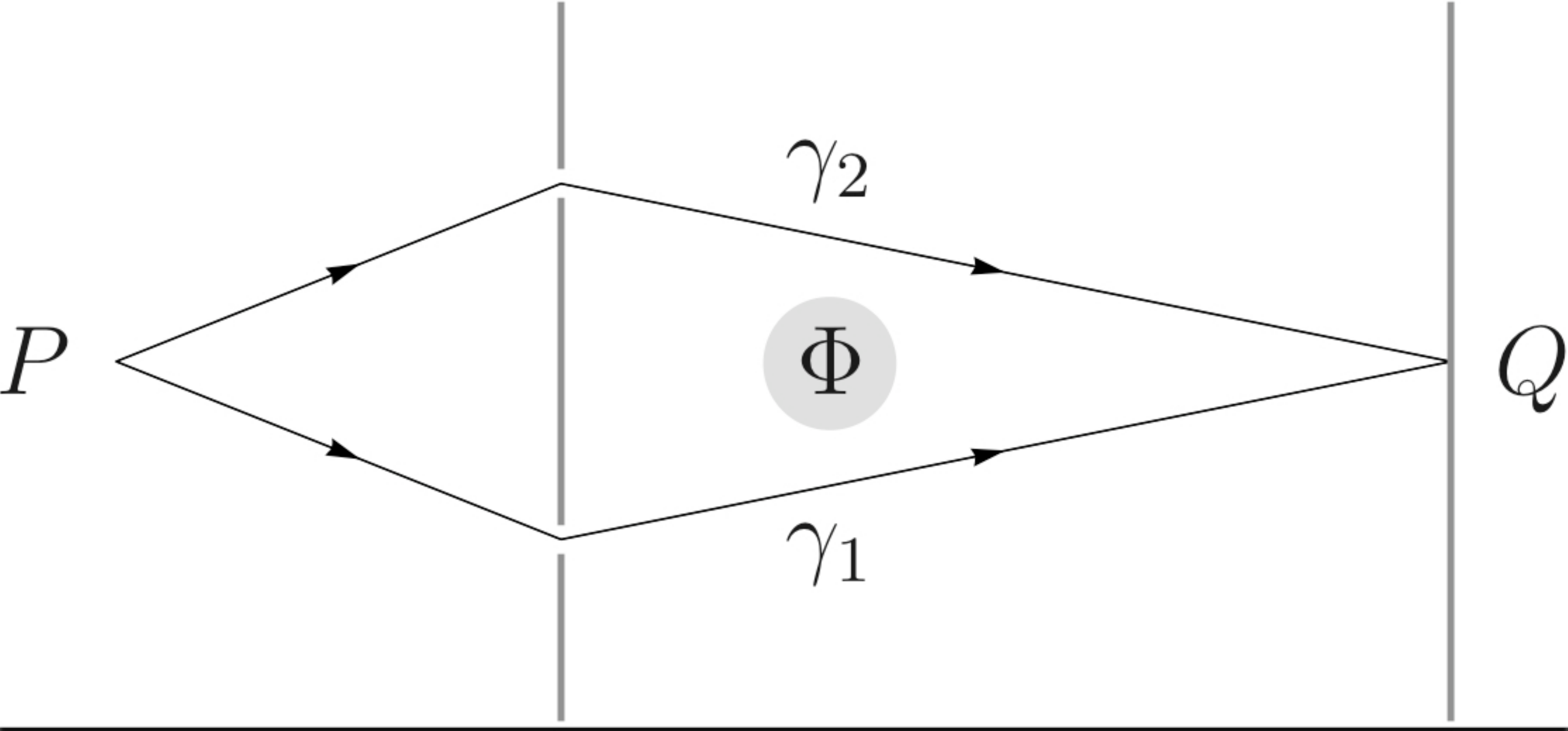}
\caption{\label{figure-AB-apparatus} A plane projection of the standard
    magnetic Aharonov-Bohm effect apparatus.}
\end{minipage}\hspace{2pc}%
\begin{minipage}{14pc}
\includegraphics[width=14pc]{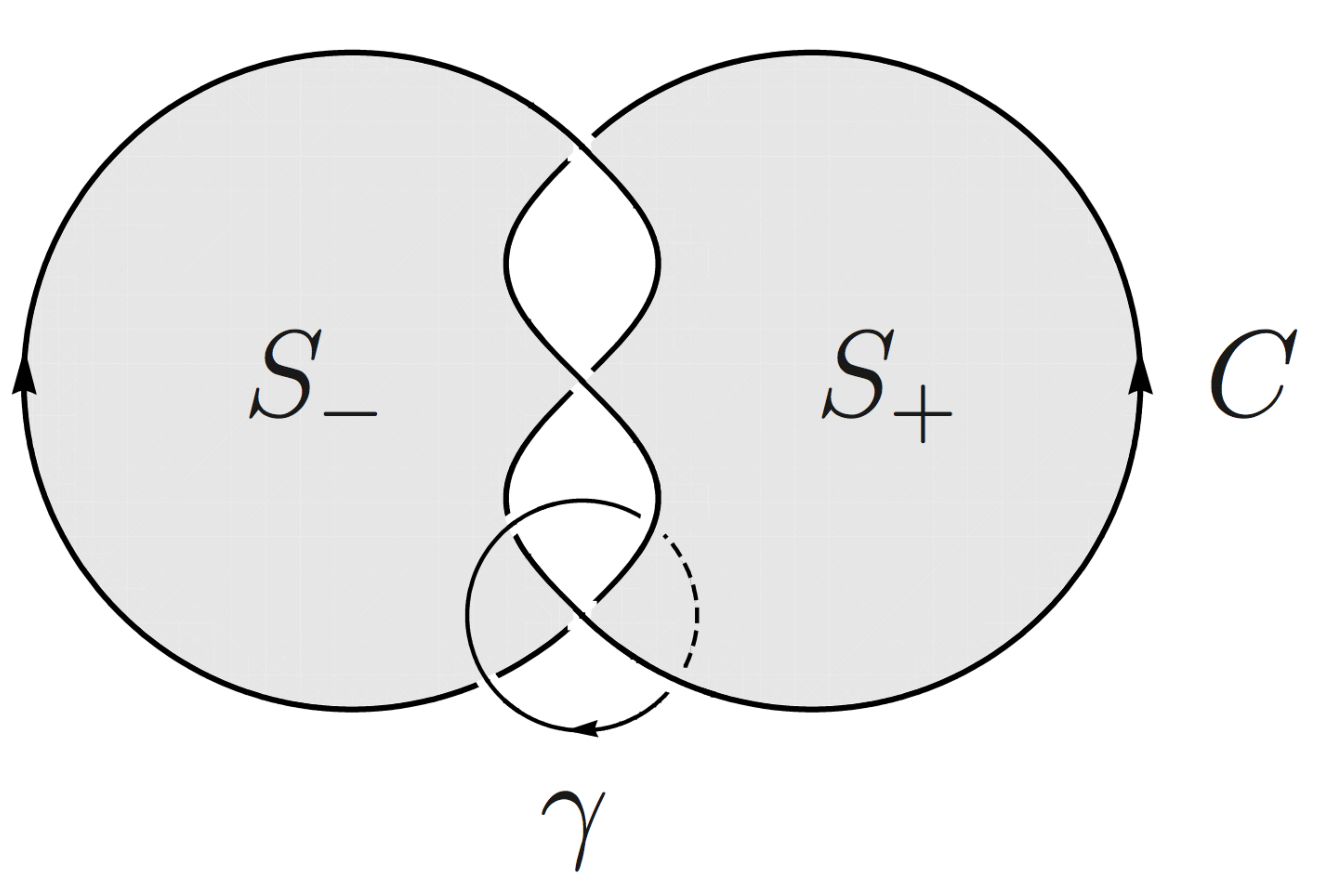}
\caption{\label{Seifert} The trefoil knot $C$ bounds an oriented
  surface $S=S_{-}+S_{+}$ called the Seifert surface. The curve
  $\gamma$ passes through a hole in the Seifert surface.}
\end{minipage} 
\end{center}
\end{figure}

In the standard Aharonov-Bohm effect, the two interfering paths of the
electron and the solenoid are both copies of $S^1$ combined to form
the Hopf link, topology of which is described by the fundamental group
$\pi_1(S^3 - S^1 ) = \mathbb{Z}.$ The resulting Gaussian linking is
distinguished by a single winding number.  Generalization of the
Aharonov-Bohm effect to a trefoil knotted
solenoids~\cite{Buniy:2008cg}, where the fundamental group $\pi_1(S^3
- K) = \mathbb{Z}*\mathbb{Z}$ has a relation $\alpha^3=\beta^2$
between its generators, leads to the following results.

Consider a solenoid along path $C$ which bounds the surface
$S=S_{-}+S_{+}$ called the Seifert surface as shown in Fig. 2, and a
wave function along path $C'$, not shown in Fig. 2.  If the wave
function along $C'$ is a closed unknot that bounds the surface $D'$
(also not shown) linked with a knotted solenoid $C$ (i.e., if $C'$
punches through $S$), then
\begin{align*}
  \phi=\oint_{C'=\partial D'}A\cdot dx=\int_{D'} B\cdot dS =
  \frac{e\Phi}{\hbar c}.
\end{align*}
On the other hand if the wave function is knotted and the solenoid is
an unknot we find (interchanging paths $C$ and $C'$)
\begin{align*}
  \phi=\oint_{C=\partial S}A\cdot dx=\int_{S} B\cdot dS =
  \frac{e\Phi}{\hbar c}.
\end{align*}
In both cases the Gaussian linking is determined by the number of
times wave function path passes through oriented surface that bounds
the solenoid.  Note that if the wave function is linked through one of
the holes in the Siefert surface, see path $\gamma$ in Fig. 2, then
there is no Gaussian linking, and no standard Aharonov-Bohm effect,
since there is no Gaussian linking.  But the curves are nonetheless
linked via higher order linking.  It is rather difficult to explore
higher order linking involving knots, so we will investigate a more
tractable case.  The simplest example of higher order linking is given
by the Borromean rings. A generalization of the Aharonov-Bohm
experiment of this type is shown in Fig. 3. After a careful choice of
an appropriate gauge, one finds that the interference in the wave
function along $C_3$ is proportional to the product of the fluxes in
the two solenoids along $C_1$ and $C_2$,
\begin{align*}
  \phi=\frac{e^2\Phi_1\Phi_2}{\hbar^2 c^2}.
\end{align*}

The analysis of The Josephson effect~\cite{josephson} has a related
generalization~\cite{Buniy:2008ch}, shown in Fig. 4, which results in
the current maximum
\begin{align*}
  I=I_0\lvert\cos{(2\pi^2\Phi_1\Phi_2/\Phi_0^2)}\rvert 
\end{align*}
between points $P$ and $Q$ in the figure that is again proportional to
the product of the fluxes in the two solenoids, where $\Phi_0$ is the
fluxoid.

\begin{figure}[h]
\begin{center}
\begin{minipage}{14pc}
\includegraphics[width=14pc]{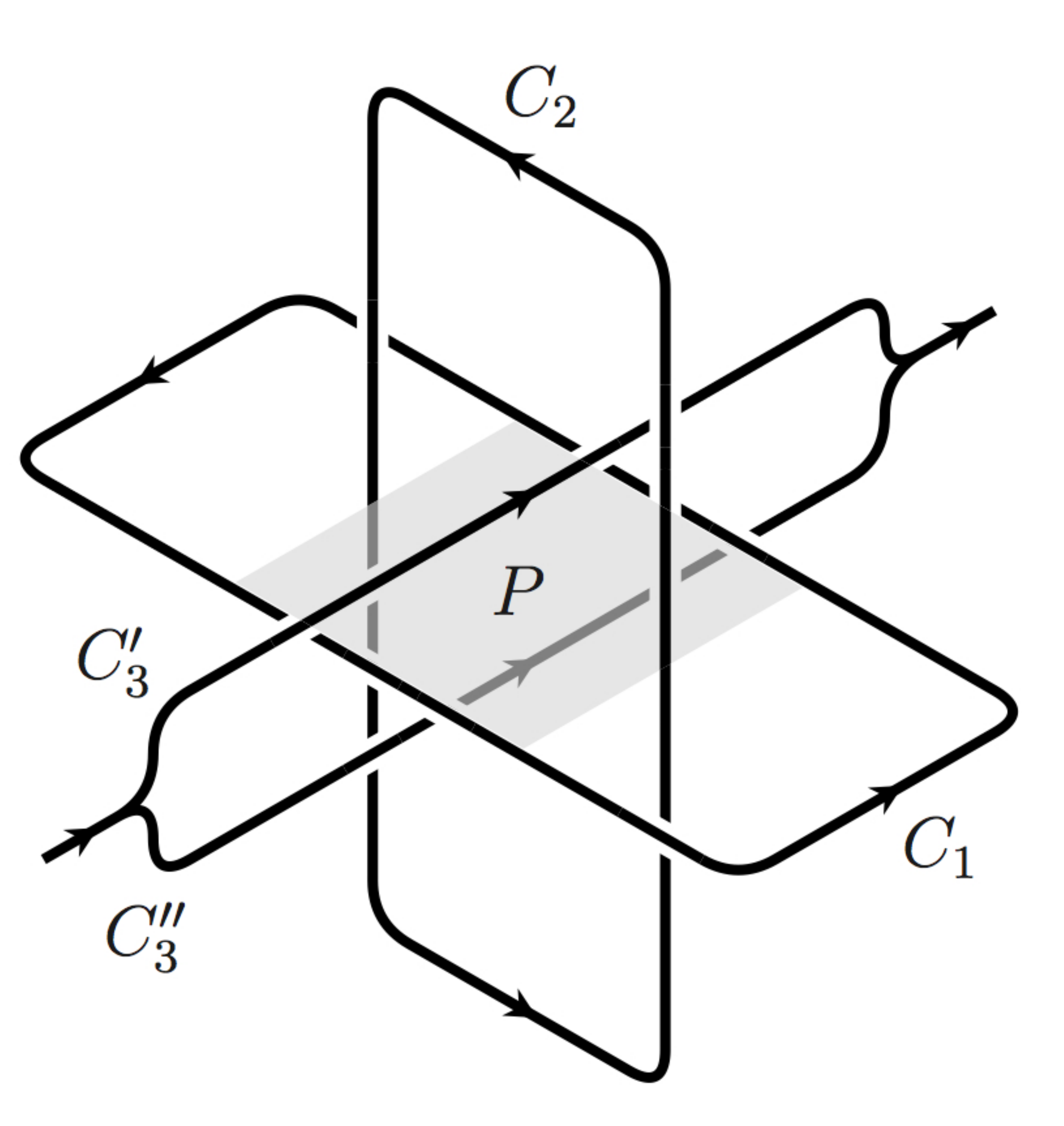}
\caption{\label{label}Schematic of a Borromean ring
    arrangement to detect the second order phase.}
\end{minipage}\hspace{2pc}%
\begin{minipage}{14pc}
\includegraphics[width=14pc]{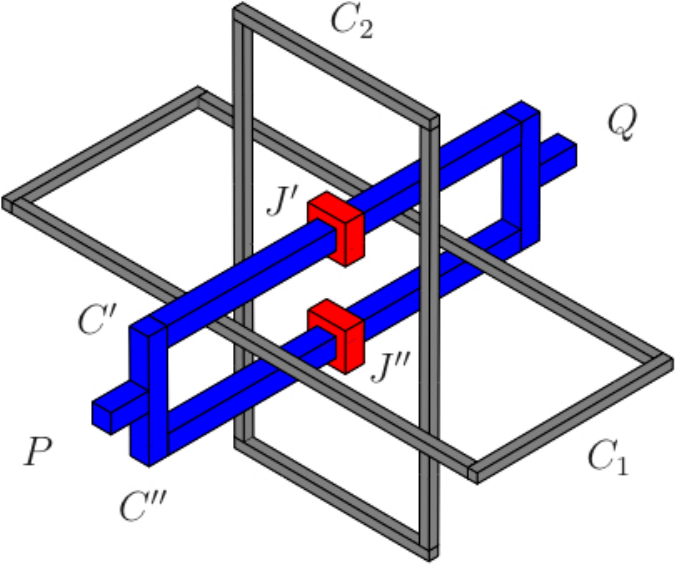}
\caption{\label{label}A schematic of a generalized magnetic Josephson
  effect apparatus. $J'$ and $J''$ are Josephson junctions. $C_1$ and
  $C_2$ are solenoids.}
\end{minipage}
\end{center} 
\end{figure}

\section{Knots and links in quantum systems}

There as many examples of quantum systems that support flux tubes,
vortices, or other tube or line like structures. For instance, such
systems include non-abelian gauge theories including quantum
chromodynamics, superconductors, superfluids, liquid crystals, and
atomic condensates to name a few.

Knots were first discussed in the context of particle physics by
Lord Kelvin who modeled ÒelementraryÓ atoms as 
knotted fluid vortices in the aether. While this model did not succeed
due to insufficient scientific knowledge at the time, it did show that Kelvin 
was ahead of his time.

\section{Tightly knotted QCD flux tubes as glueballs}

\begin{figure} 
  \includegraphics[width=15cm]{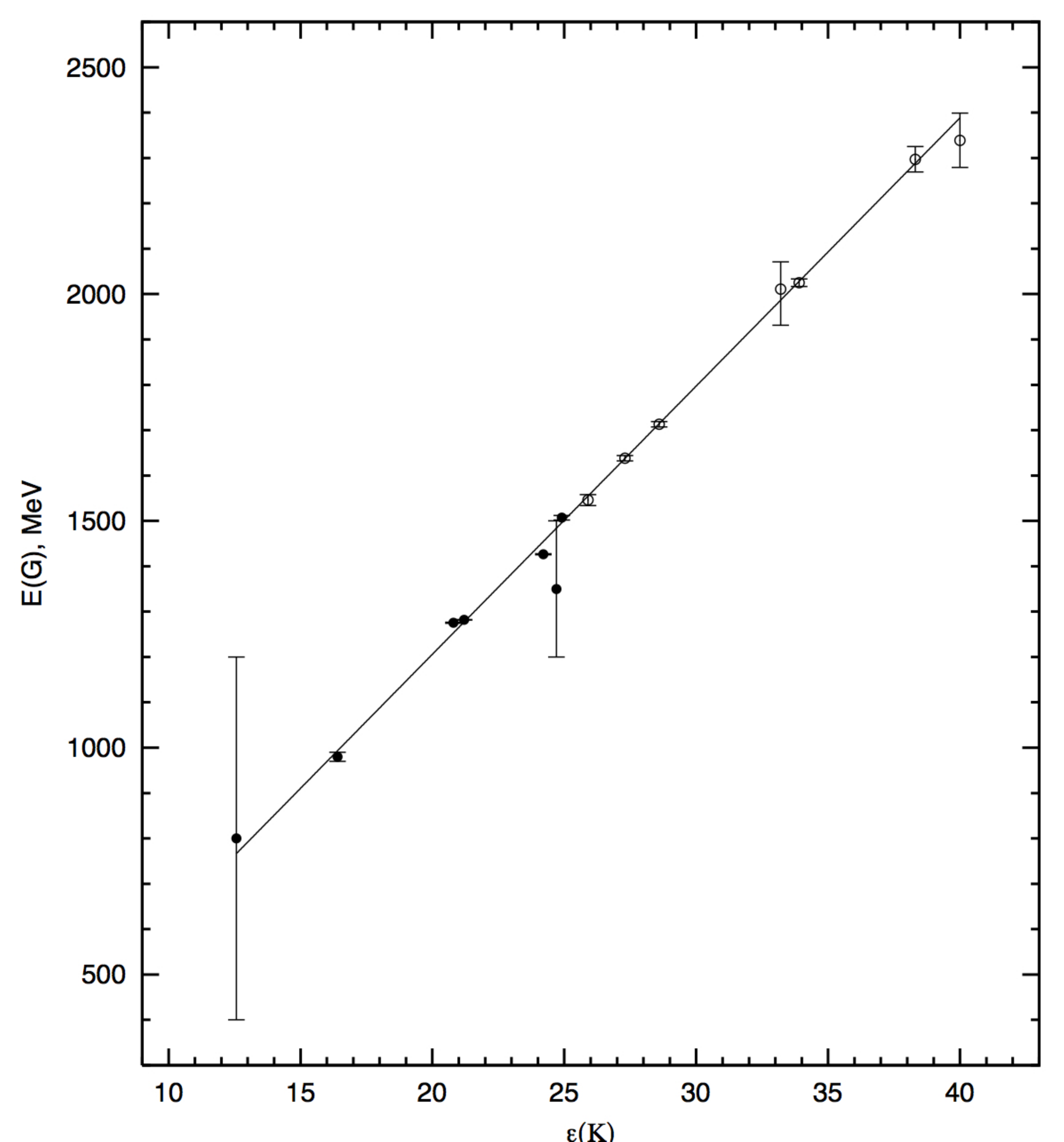}
  \caption{\label{GlueballFit.eps} Relationship between the glueball
    spectrum E(G) and ratio of knot length to flux tube radius. Each
    point in this figure represents a glueball identified with a knot
    or link. The straight line is the model fit of $f_J$ states to the
    tight knot/link spectrum from \cite{Buniy:2002yx}, which contains more discussion.}
\end{figure}

The quark model is sufficient to describe most of the spectrum of
hadronic bound states, but after filling the multiplets, a number of
states remain and it has been suggested that at least one and perhaps
several of these states are glueballs---states with no valance
quarks. Two of us have suggested that the glueball spectrum of QCD is
a result of tight knots and links of quantized chromo-electric
flux~\cite{Buniy:2002yx}.  This provides an infinite spectrum, up to
stability, of new hadronic states, and predicts their
energies. Identifying knot lengths with particle energies means the
glueball spectrum is the same as the tight knot spectrum up to an
overall scaling parameter.  A fit matching knot/link
lengths~\cite{ACPR} with the presumed $f_J$ glueball
states~\cite{Amsler:2008zz} states, is shown in
Fig.~\ref{GlueballFit.eps}, see Ref.~\cite{Buniy:2002yx}, where the
lightest $f_J$ state is identified with the shortest knot or link, the
next lightest state with the next shortest knot or link, etc.  E.g.,
we identify the $f_0(600)$ with the Hopf link, the $f_0(980)$ with the
trefoil knot, etc.  The fact that this is a one parameter fit means
that any system of tightly knotted tubes will show the same universal
behavior, with their energy/mass spectra corresponding to the length
of knots and links up to one overall scaling parameter per
system~\cite{Buniy:2002yx}.

In the present model, glueball candidates of non-zero angular
momentum, called $f_J$ states, correspond to spinning knots and
links. (These solitonic objects could also have intrinsic angular
momentum, but that will not be discussed here.)  To calculate
rotational energies we need the moment of inertia tensor for each
tight knot and link. We can get exact results for links with planar
components. E.g., in its center of mass frame, the moment of inertia
tensor for a Hopf link of uniform density is
\begin{align*} 
  {I_{\textrm{\scriptsize Hopf}}} = \left( \begin{array}{ccc} 21 & 0 &
    0 \\ 0 & \frac{75}{2} & 0 \\ 0 & 0 & \frac{75}{2}
  \end{array} \right)\pi^{2} \rho \;a^5,
\end{align*}
where one torus is in the $xz$-plane and the other is in the
$xz$-plane. Note this inertia tensor corresponds to a prolate spheroid
as do other straight chains of links with an even number of elements.
Odd straight chains do not have so much symmetry. Moment of inertia
tensors of other knots and links can be calculated by Monte Carlo
methods~\cite{ACPR}. For further discussion and a detailed analysis of
both the zero and non-zero angular momentum $f_J$ states including
calculations of moment of inertia tensor see Ref.~\cite{MJHthesis}.\\

\section{Conclusions} 

There are many classical systems that can be knotted or linked.  Some
of these knots can be tight, e.g., Taylor states in plasmas, but since
the tubes can carry an arbitrary amount of flux, their radii are not
fixed.  In contrast, the flux in quantum systems is quantized leading
to fixed radius tubes and hence quantized lengths for tight knots an
links. This in turn leads to a quantized mass/energy spectrum. More
generally, any quantum systems that supports quantized flux will have
this same universal spectrum, i.e., one parameter per system---the
slope.

\section*{Acknowledgments}

RVB acknowledges support from a DoE grant at ASU and from the Arizona
State Foundation. The work of MJH and TWK was supported by DoE grant
number DE-FG05-85ER40226.

\end{document}